\documentstyle[12pt]{article}

\begin{document}
\title {HIGGS FIELDS IN SUPERSYMMETRICAL THEORY}
\baselineskip 6mm
\author{Andrei M. Sukhov\footnotemark}%\addtocounter{footnote}{1}
\footnotetext{sukhov@mb.ssau.samara.ru}%
\date{\today}
\maketitle
{\em Department of Physics,
Samara State Aerospace University, Moskowskoe shosse 34, Samara, 443086,
Russia}
\begin{abstract}
The theory including interaction between Siegel and gauge multiplets
leads to the model of nonbreaking supersymmetry which contains
massive scalar, fourcomponent fermion and gauge fields.
The upper bound of Higgs boson mass is estimated as heavest fermion
mass.
The idea to replace Higgs field by scalar superpartner or
by auxiliary fields of corresponding
supermultiplet is discussed.

%12.60Fr; 12.15Rr
\end{abstract}
%\pacs{12.60Fr; 12.15Rr}
\pagebreak
% body of paper here
\vskip 0.5in

Realistic supersymmetry model has to contain a mechanism for the
breakdown of supersymmetry that splits masses of the different
members of supermultiplets and in addition induces the scale of the
weak interaction breakdown. One more reason why supersymmetry is of
considerable interest is that one could solve in principle Higgs
bosons problem. It is well--known that standard model is not the
final theory of the world since it contains some twenty free
parameters. There is especially one sector in the theory, Higgs
sector, which remains rather mysterious. But well-known
breaking mechanisms \cite{fay,raf,fil} say us nothing about
standard model action for a massive gauge field
because any term coupling matter fields and gauge
fields is absent in their superpotentials of interaction. It is
enough difficult to mix auxiliary and physical components thereby to
break supersymmetry.

Matter fields are considered to be described by Wess-Zumino
supermultiplet $G$ just as gauge field is included in vector
multiplet $V$ \cite{west}. W. Siegel \cite{sig} has suggested a
term of interaction between both $V$ and $G$ multiplets mentioned
above:
\begin{equation}
S_{int}=\int d^{4} x d^{4}\theta GV\sim
\int d^{4}x d^{2}\theta \psi_{\alpha}W^{\alpha} +h.c.
\end{equation}
Here $\psi_{\alpha}$ is Siegel chiral spinor superfield \cite{sig}
and $W_{\alpha}= \bar{D}^{2}D_{\alpha}V$ is field strength of
abelian vector multiplet $V$.  Such an action describes a massive
vector multiplet by gauge invariant term
($\psi_{\alpha}W^{\alpha}$).

On the other hand, Siegel multiplet can be written as linear
multiplet $(C,\chi,v_{\mu})$ which is also alternative description
of field content that are contained in Wess-Zumino model
\cite{west}. This multiplet $\psi$ may be generated from scalar
supermultiplet  $\phi$ by putting the following coupling
\begin{equation}
{\cal D}^{\scriptscriptstyle A}
{\cal D}_{\scriptscriptstyle A}\phi=0,
\end{equation}
and in fourcomponent notations the transformation laws are
\begin{equation}
\delta C=i\bar{\varepsilon}\gamma_{5}\chi,\;\delta\chi=(-i\gamma_{5}
\hat{\partial}C+\hat{A})\varepsilon,\;\delta A_{\mu}=-\bar
{\varepsilon}\sigma_{\mu\nu}\partial^{\nu}\chi.
\end{equation}
Using the rules of supertensor calculus from West textbook
\cite{west},
the action (1) may be rewritten via component fields of $G$ and of
$V$ as
\begin{equation}
S_{int}=\int d^{4}x [GV]_{D}=\int d^4 x (CD-v_{\mu}A^{\mu}- \bar
{\lambda}\chi),
\end{equation}
where $V=(A_{\mu}, \lambda, D)$.

It should be noted
that the role of Bose components of $G$ is reversed: the scalar $C$
is now physical (instead of auxiliary $D$), and the transverse
vector $v_{\mu}$ has got one physical and two auxiliary
components (instead of two physical and one auxiliary ones for
$A_{\mu}$), i.~e. in action (4) physical fields and auxiliary
fields are mixed in nontrivial way. If supersymmetry are broken
then auxiliary fields including of $v_{\mu}$--components 
get their nonvanishing vacuum expectation values.  Second term of
action (4) helps to rise a question: could we use them to clear up
Higgs boson role in standard model?

Gremmer and Scherk \cite{CrSch} tried earlier to understand how the
spontaneous symmetry breakdown can be generated without introducing
the scalar Higgs field with help of fully massless theories.
However, all calculations in papers \cite{sig,CrSch} are carried out
with abelian Lagrangian but for construction of standard model more
wide class of symmetries is used.
Therefore an attempt is made to generalize a
mechanism from Cremmer's and Scherk's article on
non--abelian multiplets.

Let us consider the Siegel multiplet $\psi_{\alpha}$ which are
transformed under non-abelian gauge group:
\begin{equation}
\psi^{a}_{\alpha}\rightarrow \psi'^{a}_{\alpha}=(e^{\Lambda})
{}^{a}{}_{b}\psi^{b},
\end{equation}
where $\Lambda^a{}_b=\Lambda^i(T_i)^a{}_b$ and $(T_i)^a{}_b$ are
generators
of gauge group. The transformation laws for $\psi'^{\alpha}$ fields
copy directly the Eq.(3) with replacing the derivative on a covariant
one.

This fields are coupled to Yang--Mills supermultiplet $V^{a}$ which
supersymmetric and gauge transformations have standard expansion
\cite{sig}.

Corresponding action is the following sum:
\begin{equation}
S=S_{\scriptscriptstyle YM}+S_{Sieg}+S_{int}.
\end{equation}
Here $S_{\scriptscriptstyle YM}$ describes Yang--Mills supermultiplet
\begin{equation}
S_{\scriptscriptstyle YM}=\int d^4 x\;Tr\left\{-\frac{1}{4}
F^a_{\mu\nu}F^{a\mu\nu}-
\frac{1}{2}\bar{\lambda^a}\hat{D}\lambda^a+\frac{1}{2}
(D^a)^{2}\right\},
\end{equation}
$S_{Sieg}$ is an action of Siegel superfield \cite{west}
\begin{equation}
S_{Sieg}=\int d^{4} x[G^aG^a]_{D}= \int d^{4} x\;Tr\left\{-\frac{1}
{2}(D_{\mu}C^a)^2 -\frac{1}{2}\bar{\chi}{}^a\hat{
D}\chi^a-\frac{1}{2} (v^a)^{2}_{\mu}\right\},
\end{equation}
and $S_{int}$ is a term of the interaction
\begin{equation}
S_{int}=\frac{k}{2}\int d^4 x[G^aV^a]_D=\frac{k}{2}\int d^4 x\; Tr
\left\{C^aD^a-A^a_{\mu}v^{a\mu}-\bar{\lambda}{}^a\chi^a\right\}.
\end{equation}
The boson part of new Lagrangian
\begin{equation}
{\cal L}_b=-\frac{1}{4}F_{\mu\nu}^{a}F^{a\mu\nu}+\frac{1}{2}(D^{a})
^2-\frac{1}{2}(D_{\mu}C^a)^2-\frac{1}{2}(v^a_{\mu})^2+
\frac{k}{2}C^a D^a-\frac{k}{2}A_{\mu}^a v^{a\mu},
\end{equation}
and the fermion part
\begin{equation}
L_f=-\frac{1}{2}\bar{\lambda}^a\hat{D}\lambda^a-\frac{1}{2}\bar
{\chi}^a\hat{D}\chi^a-\frac{k}{2}\bar{\lambda}^a\chi^a
\end{equation}
invariant under non--abelian group of the internal symmetry.

We can now diagonalize the Lagrangian (9) by rewriting it in terms of
new fields, as it has made in Gremmer's and Scherk's paper
\cite{CrSch},
\begin{eqnarray}
h_{\mu}^a&=&v_{\mu}^a+\frac{k}{2}A_{\mu}^a,\hspace{2mm}E^{a}=D^a+
\frac{k}{2}C^a,  \\
{\cal L}_b&=&-\frac{1}{4}F_{\mu\nu}^{a}F^{a\mu\nu}+
\frac{k^2}{8}(A_{\mu}^a)^2
-\frac{1}{2}(D_{\mu}C^a)^2-
\frac{k^2}{8}(C^a)^2
-(h^a_{\mu})^2 +(E^a)^2.
\end{eqnarray}
However one must be careful, because $v_{\mu}$ has one physical
degree of freedom and two auxiliary degrees of freedom and is
constrained by the
condition $D^{\mu}v_{\mu}^a=0$. Therefore  a new scalar
field $B^a$ and ortonormal basis of vectors $\mbox{\boldmath $e$}_
{\scriptscriptstyle A}$
should be introduced where $v_{\mu}^a= D_{\mu}B^a+$
two additional terms $e^{\mu}_{\scriptscriptstyle E}E^a$, $e^{\mu}_
{\scriptscriptstyle F}F^a$
for off--shell Lagrangian so that
\begin{equation}
(v_{\mu}^{a})^2=(D_{\mu}B^{a})^2+(E^{a})^2+(F^{a})^2,
\end{equation}
where $E^a$ and $F^a$ are scalar and pseudoscalar auxiliary
fields respectively.

It should be noted that Cremmer's and Scherk's conditions $E^a=0$,
$h^a_{\mu}=0$ find now a physical sense: these are the equations for
auxiliary
fields. It is nontrivial that in final expressions of equations
\begin{equation}
E^a=-\frac{k}{2}e^{\mu}_{\scriptscriptstyle E}A^a_{\mu};\hspace{2mm}
F^a=-\frac{k}{2}e^{\mu}_{\scriptscriptstyle F}A^a_{\mu};\hspace{2mm}
D^a= -\frac{k}{2}C^a,
\end{equation}
the nonzero spin fields are included.

Let us return  to fermion sector and examine the particle spectrum
more
closely. It is important to note that Majorana fermions cannot carry
any conserved additive quantum number. The charged fermions
is described by four--component Dirac spinors therefore two Majorana
states $\lambda^a$, $\chi^a$ should be mixed so that they lead to new
massive fields. Such a model suggested by Wolfenstein \cite{h}
is used often for a description of the neutrino oscillation.
This choice is known to correspond a very general
massive term of the four--component fermion field \cite{chl}.
It is convenient to choose new spinors $\eta,\;\omega$ obtained
after rotation in a plane $O(\lambda,\chi)$. Here
\begin{eqnarray}
\eta&=&\lambda\cos \theta -\chi\sin\theta=\psi_{\scriptscriptstyle L}
+ \psi_{\scriptscriptstyle L}^c,\\
\omega&=&\lambda\sin\theta+\chi\cos\theta=\psi_{\scriptscriptstyle R}
+ \psi_{\scriptscriptstyle R}^c
\end{eqnarray}
are self--conjugative fields with respect to charge--conjugation
operator $C$. The new mass term \cite{chl} is equivalent to 
\begin{equation}
L_{fm}=\frac{k}{2}(\bar{\eta}\omega\cos^2\theta +\bar{\omega}
\eta\sin^2\theta)+\frac{k}{4}(\bar{\eta}\eta+\bar{\omega}\omega)
\sin {2\theta}.
\end{equation}
It means that fermion sector $\psi^a$ of standard model including neutrinos
$\nu_l$
and charged leptons $l$ may be wholly described with help of such a set
and of the transformations of the internal symmetry group, i.e.
\begin{equation}
L_{fm}=\frac{k}{2}\bar{\psi}^a\psi^a+A\bar{\psi}^{ac}
_{\scriptscriptstyle L}
\psi^{a}_{\scriptscriptstyle L}+
B\bar{\psi}^{ac}_{\scriptscriptstyle R}
\psi^{a}_{\scriptscriptstyle R},
\end{equation}
where $\psi^a$ is the fields implemented the representation of symmetry
group, for example, $\psi^a=(e_{\scriptscriptstyle L},\nu_e)$.
Left or right components of $\psi$ could be included in
supermultiplets $G$ or $V$ correspondingly.

Thus, our model of nonbreaking supersymmetry
contains massive scalar field $D^a$, fermion family $\psi^a$ and gauge
field $A_{\mu}^a$ transformed under internal symmetry group.
Here, $v^a_{\mu}$ is simply the Goldstone bosons which are eaten by
gauge fields $A_{\mu}^a$ thereby giving mass to the supermultiplet
members.

The question remains now whether the breakdown of internal symmetry
can be  related to the breakdown of supersymmetry.
The first of the possible scenarios is that to break
internal symmetry at low energy. Then, the vacuum expectation
values of all auxiliary fields vanish
\begin{equation}
\langle E^a \rangle=
\langle F^a \rangle=
\langle D^a \rangle=0
\end{equation}
and the supersymmetry is unbroken.
This model contains only scalar $C^a$ from new (nondetectored)
particles and dynamical term $\frac{1}{2}(D_{\mu}C^a)^2$
in Lagrangian (13) allows to choose it
on a role of Higgs field to provide standard model with
soft breakdown of the internal symmetry.

It should be noted the absent of
Yukawa terms generating the fermion masses. These
masses as well as Higgs boson mass are determined by Siegel
interaction, see Eqs.(13,18). For construction of reach picture
of fermion and of
boson sectors the transformations of the internal symmetry have to be
engaged in addition to mixed mechanism mentioned above. An upper
bound of Higgs mass is determined now by mass
$\frac{k}{2}$ of Dirac fermion in Eq.(19), the valueof this mass 
is probably a few {\em GEV}.

Besides, present theory prompts us to consider the production of
gauge bosons as supersymmetrical transition inside of
Yang--Mills multiplet $V^a$ ($\lambda^a\rightarrow A^a_{\mu}$).
Therefore, Higgs field presented by scalar $C^a$  could be
an superpartner of right--handed
component of Dirac state.

Here $L-R$ symmetry is conserved and Dirac field $\psi_a$ can carry 
additive quantum number, e.g. electric charge. For neutrino Majorana
mass term violates only lepton family number.  

Consequently, the Higgs does not couple directly to left-handed components
of quarks and leptons and its
production and detection in experiment remains difficult problem. An
exception is production in the decay of $Z$ boson, {\em i.e.}
$e^{-}e^{+}\rightarrow Z^{*}\rightarrow ZH$ process.

Another
possibility of theory construction is opened when masses of
superpartners $\frac{k}{2}$ coincide with mass of gauge bosons
$W^{\pm},Z$.
Then the idea to replace Higgs fields by corresponding auxiliary
fields $E^a,F^a,D^a$ of $G$ and $V$ supermultiplets may be applied.
Indeed Lagrangian of standard model after symmetry breaking contains
both nonvanishing vacuum expectation value of Higgs field and
physical component or so-called Higgs boson. In order to obtain
on--shell supersymmetric Lagrangian we have only to change auxiliary
fields $E^a, F^a, D^a$ on their vacuum expectation values
$\langle E^a \rangle,
\langle F^a \rangle,
\langle D^a \rangle$.
If auxiliary
fields is chosen on role of Higgs fields then after supersymmetry
breaking corresponding Lagrangian does not include any additional
physical state in comparison with initial off--shell Lagrangian.
As former, transformation of internal symmetry has to take a part in
construction of final theory.

Besides, the realistic model must
contain a mechanism for the breakdown of supersymmetry that splits
the masses of different members of the supermultiplet and in addition
induces the scale of the breakdown of internal symmetry.
Up to now there is no model \cite{HabK} which is
theoretically completely satisfactory in this respect.  It seems
however that these questions can be now answered at energy scales
less then $M_{\scriptscriptstyle W}\sim$ 100 $GeV$ and it is necessary
to investigate right--handed sector of supersymmetrical models.

\pagebreak


\begin{thebibliography}{9}
\bibitem{fay}
P. Fayet, Phys. Let., {\bf 58B} (1975) 67
\bibitem{raf}
L. O'Raifeartaigh, Nucl. Phys. {\bf B96} (1975) 331
\bibitem{fil}
P. Fayet and J. Illiopoulos, Phys. Let. {\bf 51B} 461
\bibitem{west}
P. West, {\em Introduction to supersymmetry and supergravity} (World
Scientific, 1986)
\bibitem{sig}
W. Siegel, Phys. Let. {\bf 85B} (1979) 333
\bibitem{CrSch}
E. Cremmer and J. Scherk, Nucl. Phys., {\bf B72} (1974) 117
\bibitem{h} L.
Wolfenstein, Nucl.  Phys.,{\bf B186} (1981), 147.
\bibitem{chl}
T.P. Cheng and L.F.Li, Phys. Rev., {\bf D22}, (1980) 2860
\bibitem{HabK}
H.E. Haber and G.L. Kane, Phys. Rep. {\bf 117} (1985) 77
\end{thebibliography}
\end{document}